cond-mat/9212004

\documentstyle[12pt]{article}

\begin{document}
\begin{center}
\begin{large}
{\bf Fractal Properties of the Distribution of Earthquake Hypocenters}
\end{large}

\bigskip
\bigskip

Hisao Nakanishi$^{\dag}$ and Muhammad Sahimi$^{\ddag}$\\
HLRZ, KFA--J\"{u}lich, Postfach 1913\\
W-5170 J\"{u}lich, Germany

\bigskip

Michelle C. Robertson and Charles C. Sammis\\
Department of Geological Sciences, University of Southern California\\
Los Angeles, CA 90089, U.S.A.

\bigskip

Mark D. Rintoul\\
Department of Physics, Purdue University\\
W. Lafayette, IN  47907 U.S.A.
\end{center}

\bigskip

\begin{flushleft}
$^{\dag}$Present and permanent address: Department of Physics,
Purdue University,\\West Lafayette, Indiana 47907 U.S.A.\\
$^{\ddag}$Present and permanent address: Department of
Chemical Engineering,\\University of Southern California,
Los Angeles, CA 90089-1211 U.S.A.
\end{flushleft}

\bigskip

We investigate a recent suggestion that the spatial distribution of
earthquake hypocenters makes a fractal set with a structure and
fractal dimensionality
close to those of the backbone of critical percolation clusters, by
analyzing four different sets of data for the hypocenter distributions
and calculating the dynamical properties of the geometrical distribution
such as the spectral dimension $d_s$.  We find that the value of $d_s$ is
consistent with that of the backbone, thus
supporting further the identification of the hypocenter distribution
as having the structure of the percolation backbone.

\bigskip
\bigskip


\newpage

Two important and related phenomena in natural rock masses are
earthquakes and the nucleation and propagation of fractures.
Earthquakes are the result of a series of complex phenomena involving
the interaction between stress concentration and fluid flow, and have
been the subject of considerable recent interest
[1--7]. They have been proposed [3--5] to be related to self-organized
critical phenomena (SOCP), in that they are the product of a dynamical
many-body system that reaches a critical state without fine-tuning its
parameters. Such systems reach a stationary critical state which is
characterized by spatial and temporal correlations that follow
power-laws without any intrinsic length or time scales.
Although the connection between earthquakes and SOCP has led to a
deeper understanding of earthquake phenomenon, a clear geometrical
interpretation of the spatial distribution of earthquakes, which is
essential for constructing realistic spatial and
temporal correlation functions for earthquakes, was lacking
until recently. On the other hand, most natural rock
masses contain large fractures, in the form of a complex and
interconnected network \cite{mo1}, the presence of which is crucial
to the higher production of oil from underground reservoirs, generation
of heat and vapor from geothermal reservoirs, and the development of
groundwater resources. It had been argued \cite{kagan} that the
spatial distribution of earthquakes is closely related
to the structure of fracture networks in rocks. However, this connection
had not been quantitatively established.

In a recent paper \cite{mo2} a quantitative connection was proposed
between the structure of fracture networks and the spatial distribution
of earthquakes.  Analyzing extensive geological data and using computer
simulation models of fracture \cite{mo3}, it was proposed that
{\em large scale} fracture networks (order of kilometers) of
heterogeneous rocks are fractal sets whose structures are similar
to critical percolation clusters with a fractal dimensionality
\cite{feder} $d_f \simeq 2.5$. Moreover, since earthquake hypocenters
are usually on fracture and fault networks of rocks, they have to
belong to the {\em active} part of the networks where large scale
deformations and stress transmission take place, i.e., earthquake
hypocenters have to belong to the {\em backbone} of fracture networks.
Indeed, the analysis \cite{mo2} of four different sets of data for the spatial
distribution of earthquake hypocenters indicated that the centers are on a
fractal set with a fractal dimensionality $d_f \simeq 1.8$, close to that
of the backbone \cite{stauffer}
of three-dimensional critical percolation cluster.

Although the closeness of the fractal dimensionalities suggests
the connection between the earthquake hypocenter distribution and the
percolation backbone, it is not entirely conclusive, since two
fractal sets may have the same fractal dimensionality but rather
different structures. In this article we further explore
this connection by calculating some {\em dynamical} properties of the
fractal structures of the earthquake hypocenter distribution and by
comparing them with those of the percolation backbone.

We have taken four seismic data sets from four different regions in
Southern California, namely, San Andreas--Elsinore (SA--EL),
Parkfield, Whittier, and Upland (see Table~\ref{data}).
Two of these are the same ones as
analyzed in \cite{mo2} for the fractal dimensionality $d_f$
and the other two are also essentially the same except for the
minor additions of data points.  In order to test
the hypothesis that the distribution
of hypocenters forms a structure similar to that of, say, the lattice
nodes contained in a percolation backbone created on that lattice, what
we need to do first is to place the earthquake centers on a fictitious
lattice network.  Then we can investigate the dynamical (and static)
properties of this connected network, the nodes of which are the earthquake
hypocenters.  Physically, we would expect this fictitious network
to correspond in some way to the actual fracture network on which the
earthquake locations must lie; however, for the present analysis, it
is immaterial whether there is a direct correspondence between the
supposed connected network and the actual fractures in the rocks.

Thus, we first transform the hypocenter distribution data by a form of
coarse graining; i.e., the data points expressed in $(x,y,z)$ coordinates
in integer units of $100m$ are linearly scaled to bring the outlying
points closer and in the process those points that fall within
a given distance from each other are replaced by a single point.
Arriving at integer coordinates in units of arbitrary lattice constant
(in all cases within the volume of $50\times 50\times 50$),
we then overlay the
connectivity of a simple cubic lattice, either only to the first neighbors
or up to further neighbors.  When this is done, the network of lattice
nodes typically breaks up into many disconnected {\em clusters}, most of
which are tiny but there is always one cluster which
is composed of the bulk of the nodes in the network.
We focus on this largest {\em connected} cluster in all cases and study
its dynamical and static properties.  Of course, we must make sure
that this transformation of the original data has not distorted their
geometrical characteristics; we will discuss some checks on this
point later on.  The parameters of this
transformation are also given in Table~\ref{data}.

By the dynamical properties, we mean the properties associated
with diffusion on the connected cluster if the cluster were used as the
channel for diffusion (or random walk).  By the mapping
between diffusion and vibration \cite{alexander}, we may equivalently
characterize this work as studying the vibrational spectrum of an
elastic network having the same geometric structure as our cluster.
Thus, e.g., the probability $P(t)$ of a random walk on this cluster
to return to its starting point after $t$ steps is related to the vibrational
density of states of the corresponding elastic network
by a Laplace transform \cite{alexander}.
(However, it is important to keep in mind that all of this is simply
a tool in this case to characterize the {\em geometrical} structure of
the distribution of the earthquake hypocenters and has nothing to do
with the vibration of the rocks per se.)

In terms of the random walk problem, the spectral dimension $d_s$ and
the random walk dimension $d_w$ \cite{feder,stauffer} can be defined
by the relations:
\begin{eqnarray}
P(t) & \sim & t^{-d_s /2} \\
R(t) & \sim & t^{1/d_w} ,
\end{eqnarray}
where $R(t)$ is the root-mean-square displacement of the random walk
in time $t$.  However, rather than simulating random walks to
calculate $d_s$ and $d_w$, it is in many ways simpler to perform
the spectral analysis as suggested by the vibration--diffusion
mapping.  We do this following the method of Ref. \cite{fuchs,nakanishi}
by first constructing the {\em hopping probability matrix} ${\bf W}$
where $W_{ij}$ is the probability for the random walker to hop from
site $j$ to $i$ per time step and then diagonalizing ${\bf W}$ to
obtain eigenvalues $\lambda$ and eigenvectors near the maximum
eigenvalue with high accuracy.  To be specific, we use the so-called
{\em blind ant} model of the random walk \cite{havlin}, for which
${\bf W}$ is symmetric and the diagonal terms are generally non-zero,
but the specific choice of the random walk kinetics is irrelevant
for our purposes.  Once the diagonalization is done, we compute two
quantities, the density of eigenvalues $n(\lambda )$ and a certain
function $\pi (\lambda )$ (which is the product of $n(\lambda )$ and some
coefficient determined when the stationary initial state distribution
is expanded in terms of the eigenvectors of ${\bf W}$ \cite{jacobs}).
These functions are expected to behave, asymptotically near
$\lambda =1$ \cite{nakanishi}, as
\begin{eqnarray}
n(\lambda ) & \sim & | \ln \lambda |^{d_s /2-1}
\label{dens}  \\
\pi (\lambda ) & \sim & | \ln \lambda |^{1-2/d_w} .
\end{eqnarray}

The results of fitting the transformed data to Eq.(\ref{dens})
for $d_s$ are shown in Fig.~1, where $n(\lambda )$
from two of the four data sets (SA--EL and Whittier) is plotted
against $|\ln \lambda |$ in a
double logarithmic plot and the respective linear least squares
fits are also drawn.  Clearly the data scatter fairly widely and
the exponent estimates are not expected to be very accurate.  (The
remaining two data sets have slightly greater data scatter
but with comparable slopes.)  Nonetheless, the central estimates
of the slopes from the four sets (only two are shown for clarity)
point to a value in the range of $d_s \simeq 1.18$
to $1.29$ with the fit to the largest data set by far (SA--EL)
yielding $d_s \simeq 1.19 \pm 0.13$.
On the other hand, estimating $d_w$ from these data is much
more difficult because of the much greater data scatter for
$\pi (\lambda )$.  Consequently,
we do not make numerical estimates of $d_w$ but rather only
state the result that the widely scattered data are nonetheless consistent
with the backbone values of $d_w$ in the sense that similar
fitting procedure yields exponent ranges well encompassing
the backbone value.
We summarize the exponent estimates in Table~\ref{exp} where
the error estimates are simply from the least squares fitting
and do not take into account any finite size effects or
other systematic errors that may be present.

These estimates are clearly consistent with the corresponding exponent
$d_s^B$ for the backbone of the three dimensional critical
percolation cluster: the latter can be obtained from the
scaling relation \cite{havlin}
\begin{eqnarray}
d_s^B & = & \frac{2d_f^B}{d_w^B} = \frac{2d_f^B}{2+d_f^B-d+\mu /\nu} \\
      & \simeq & 1.16 \pm 0.02  .
\label{eq:3}
\end{eqnarray}
Here the superscript $B$ denotes the backbone, $\mu$ and $\nu$
are the DC conductivity and correlation length exponents, respectively,
and the error in the numerical value is from the uncertainty
in $\mu /\nu$ ($\simeq 2.27 \pm 0.03$ \cite{pandey}) and
in $d_f^B$ ($\simeq 1.75 \pm 0.04$ \cite{herrmann}).
Although there are more recent estimates of $\mu /\nu$, e.g.,
by combining the results of Duering and Roman \cite{duering}
and of Grassberger \cite{grass}, this would give
$\mu /\nu \simeq 2.34 \pm 0.08$, consistent with but less accurate
than the result we use.  Also
a more recent estimate of $d_f^B$ \cite{danny} leads to a slightly
higher $d_s^B$ but still clearly distinguishable from the full percolation
cluster value of $d_s^P \simeq 1.328 \pm 0.006$ \cite{rammal}.
(Unfortunately, we do not know of any direct calculations of
$d_s^B$, and as is often the case with quantities obtained through
complicated scaling relations, the error given above
may be significantly underestimated.)

The transformed data form relatively small connected clusters and
thus it may be prudent to compare these results with similarly sized
{\em single} percolation backbones obtained by direct simulation on
a simple cubic lattice.
In Fig.~2, the density of states $n(\lambda )$
from three data sets are plotted for: (a) a backbone cluster of
$287$ sites, (b) a cluster of $270$ sites obtained by scaling
by a factor of $0.5$ from a larger backbone cluster of $619$ sites,
and (c) a {\em full} percolation cluster of $344$ sites (all at
the percolation threshold $p_c$).  It can be seen that in both
cases (a) and (b), the scatter of data are comparable to the
transformed earthquake data in Fig.~1 and the slopes
are also very similar.  Moreover, the data (c) has significantly
less scatter and shows a clearly different slope, corresponding
to the full percolation cluster value of $d_s$.
This figure thus shows that our transformed earthquake data have
a very similar behavior characteristic of a {\em small} backbone cluster
and that the scaling transformation used to transform the data
apparently does not affect the characteristic power-law of the
density of states.

As a further check of the possible effects of the transformation
applied to the earthquake data, we have measured the fractal
dimension $d_f$ of the earthquake data before and after
the transformation.  For this purpose, we use
the box counting method \cite{feder,fract}
where the minimum number $N(L)$ of cubes of side $L$ required
to cover the data points completely are measured.
We then obtain an estimate of $d_f$
from the relation $N(L) \sim L^{-d_f}$.
The numerical estimates of $d_f$ are given
in Table~\ref{exp}, where the fitting regions roughly
correspond to those used in Ref. \cite{mo2} for the
original data and the lower cutoffs in the transformed
data are obtained by using the scale factors of Table~\ref{data}.
{}From these, we can see clearly that the
fractal dimension of the earthquake data is not significantly
affected by the scaling and identification of a single connected
cluster from among the data points based on imposed connectivity
to either the first (for the SA--EL data) or the second neighbor
distances (for the remaining data).

It would be interesting, as a complementary task, to look for the
dependence of the behavior of $n(\lambda )$ on the cluster size for the
small critical backbone clusters.  For an asymptotically large backbone,
we obviously expect results consistent with Eq.(\ref{eq:3}).  However,
systematic tendency in finite size effects is usually observable only
when a large number of realizations of the finite size systems are
averaged. Thus, for {\em single} small clusters, the cluster to cluster
fluctuations are very large and moreover $n(\lambda )$ for an
individual cluster is not a very smooth function of $|\ln \lambda |$,
so that it is difficult to analyze for any systematic finite size
effects.  Indeed, two particular backbone clusters
of $287$ and $297$ sites (on $24^3$ grids) gave the slope of
$-0.39 \pm 0.06$ and $-0.43 \pm 0.03$, respectively, in a plot like
Fig.~2, while those of $619$ and $588$ sites (on $36^3$ grids) gave
the slope of $-0.39 \pm 0.03$ and $-0.43 \pm 0.03$, respectively.
For larger clusters of $1018$ and $1005$ sites (on $48^3$ grids),
and taking the range to include the maximum $\lambda$ after unity,
the corresponding slopes were $-0.45 \pm 0.05$  and $-0.43 \pm 0.04$,
respectively.  Since the purpose of the present work is to demonstrate
the similarity of the earthquake data and {\em single} backbone clusters
of corresponding size, we defer the systematic study of such finite size
effects to a future work.

In summary, we have presented the analysis of the dynamical
properties of the geometric network represented by the four
earthquake hypocenter distributions in Southern California.
Mainly based on the good agreement between the measured spectral
dimension $d_s$ and fractal dimension $d_f$ and
those of the critical percolation backbone
in three dimensions, we believe the case supporting the idea that
these earthquake distributions lie on the percolation backbone
has been strengthened.  Clearly, it is desirable to establish
that the physical network of active fractures is properly
represented by the connectivity we imposed in this calculation.
Although we have not done this, we believe
that this idea has sufficient supporting evidence now to deserve further
attention.

\bigskip

{\bf Acknowledgments}

\bigskip

This work was carried out while two of us (H.N. and M.S.) were
visiting the HLRZ Supercomputer Center at KFA--J\"{u}lich in Germany.
We would like to thank the Center and Hans Herrmann for warm hospitality.
In addition, M.S. is grateful to Alexander von Humboldt Foundation
for a research fellowship.

\newpage

\begin{center}
\begin{large}
{\bf FIGURES}
\end{large}
\end{center}

\bigskip
\bigskip

\begin{description}
\item[Fig. 1:]
Density of eigenvalues $n(\lambda )$ from the data sets
derived from SA--EL ($\bigcirc$) and Whittier ($\Box$) regions
are shown in double logarithmic plot against $|\ln \lambda |$.
The least square fitted lines have slopes of $-0.41 \pm 0.07$
and $-0.37 \pm 0.09$, respectively.

\bigskip

\item[Fig. 2:]
Density of eigenvalues $n_B (\lambda )$ from comparably sized
backbone and full percolation clusters at $p_c$.  The symbols
$\bigcirc$, $\Box$, and $\Delta$ correspond to a backbone
cluster of $287$ sites, a cluster of $270$ sites obtained
by linear scaling from a larger backbone cluster, and
a full percolation
cluster of $344$ sites, respectively.  The linear least squares
fitting yields lines with slopes of $-0.39 \pm 0.06$
(solid line for $\bigcirc$), $-0.40 \pm 0.11$
(dash-dotted line for $\Box$),
and $-0.36 \pm 0.03$ (dashed line for $\Delta$).
\end{description}

\newpage

\begin{center}
\begin{large}
{\bf TABLES}
\end{large}
\end{center}

\bigskip

\begin{table}[h]
\caption{Brief descriptions of the four earthquake hypocenter
distributions and the transformations used to obtain the connected
lattice clusters used in the analysis. The last column
indicates the connectivity imposed up to the indicated neighboring distance.
Each event represents an earthquake with magnitude greater than unity.}

\bigskip

\begin{tabular}{cccccc} \hline\hline
   Region & Range ($100m$) & Events & Scale Factors & Sites & Connectivity \\
\hline
   SA--EL & $934\times 1823\times 210$ & 2004 & $0.025\times 0.025\times 0.025$
& 419 & 1 \\
   Parkfield & $145\times 168\times 154$ & 885 & $0.25\times 0.25\times 0.25$ &
326 & 2 \\
   Whittier & $129\times 145\times 210$ & 224 & $0.125\times 0.125\times 0.075$
& 140 & 2 \\
   Upland & $139\times 156\times 182$ & 291 & $0.125\times 0.125\times 0.1$ &
129 & 2 \\
\hline\hline
\end{tabular}
\label{data}
\end{table}

\bigskip

\begin{table}[h]
\caption{Numerical estimates of the exponents $d_f$ for the
earthquake data before and after the scaling/connection
transformation, and that of $d_s$ for the transformed cluster.
Error estimates are only the least squares fitting errors.}

\bigskip

\begin{tabular}{cccc} \hline\hline
   Region & $d_f$ (original) & $d_f$ (transformed) & $d_s$\\
\hline
   SA--EL & $1.72 \pm 0.03$ & $1.75 \pm 0.04$ & $1.19 \pm 0.13$\\
   Parkfield & $1.76 \pm 0.03$ & $1.69 \pm 0.08$ & $1.29 \pm 0.20$\\
   Whittier & $1.73 \pm 0.03$ & $1.92 \pm 0.16$ & $1.26 \pm 0.17$\\
   Upland & $1.79 \pm 0.02$ & $1.72 \pm 0.11$ & $1.18 \pm 0.37$\\
\hline\hline
\end{tabular}
\label{exp}
\end{table}

\end{document}